\documentclass[structabstract]{aa}
\usepackage{graphicx}
\usepackage{array}
\usepackage{txfonts}
\usepackage{natbib}
\usepackage{booktabs,caption,fixltx2e}
\usepackage[flushleft]{threeparttable}
\bibpunct{(}{)}{;}{a}{}{,}

\title{Charge exchange in the ultraviolet: implication for interacting clouds in the core of NGC~1275}
\author{ Liyi Gu \inst{1} 
\and 
Junjie Mao \inst{1,2} 
\and 
Christopher P. O'Dea \inst{3,4} 
\and 
Stefi A. Baum \inst{3,5}
\and 
Missagh Mehdipour \inst{1}
\and
Jelle S. Kaastra \inst{1,2}}

\offprints{Liyi Gu (l.gu@sron.nl)}
\date{\today}

\institute{SRON Netherlands Institute for Space Research, Sorbonnelaan 2,
           3584 CA Utrecht, the Netherlands 
           \and 
Leiden Observatory, Leiden University, PO Box 9513, 2300 RA Leiden, the Netherlands 
\and 
Department of Physics and Astronomy, University of Manitoba, Winnipeg, MB R3T 2N2, Canada
\and 
School of Physics and Astronomy, Rochester Institute of Technology, Rochester, NY 14623 USA
\and 
Center for Imaging Science, Rochester Institute of Technology, Rochester, NY 14623 USA
           }
\abstract
{ Charge exchange emission is known to provide a key diagnostic to the interface between hot and cold matter in many astrophysical environments. Most of the recent charge exchange studies focus on its emission in the X-ray band, but few on the UV part, although the latter can also provide a powerful probe of the charge exchange process.}
{ An atomic calculation, as well as an application to observed data, are presented to explore and describe the potential use of the UV data for the study of cosmic charge exchange.} 
{ Using the newest charge exchange model in the SPEX code v3.03, we re-analyze an archival Hubble STIS data of the central region of NGC~1275. }
{ The NGC~1275 spectrum shows hints for three possible weak lines at about 1223.6~{\AA}, 1242.4~{\AA}, and 1244.0~{\AA}, each with a significance of about $2-3\sigma$. The putative features are best explained by charge exchange between highly ionized hydrogen, neon, and sulfur with neutral matter. The wavelengths of the charge exchange lines are found robustly with uncertainties $\leq 0.3$~{\AA}. The possible charge exchange emission shows a line-of-sight velocity offset of about $-3400$ km s$^{-1}$ with respect to the NGC~1275 nucleus, which resembles one of the Ly$\alpha$ absorbers reported in \citet{baum2005}. This indicates that the charge exchange lines might be emitted as the same position of the absorber, which could be ascribed to outflowing gas from the nucleus.} 
{}
\titlerunning{Charge exchange in the UV band}
\authorrunning{L. Gu}

\begin{document}

\maketitle

\section{Introduction}

Charge exchange (CX hereafter) is a key physical process occurring at the interface of cold and hot cosmic plasma. It is well known to produce signature line emission in the X-rays, which provide an important diagnostic to the interaction at the interface. The CX X-rays were first observed in comets as they interact with the solar wind \citep{lisse1996, cravens1997}. Followup studies also revealed CX X-rays from the geocoronal and heliospheric areas, as well as from planets in the Solar system \citep{snowden2004, dennerl2006, fujimoto2007, br2007, smith2014}. Recently, CX is further considered as a potential mechanism for X-rays from stellar winds \citep{pollock2007}, supernova remnants \citep{lallement2009, katsuda2011}, starburst galaxies \citep{tsuru2007, liu2011}, and even galaxy clusters \citep{fabian2011, gu2015, hitomi2016}. Most applications of CX to extrasolar X-ray sources still remain speculative, since the CX X-ray lines from those objects are often found to be rather weak. 

The study of CX as an astrophysical process has a long history and is not restricted to X-rays. In the planetary nebula NGC~7027, the ionization structure
observed with optical spectra can be explained by including CX excitation with hydrogen atoms \citep{peguignot1978}. \citet{chevalier1980} reported CX excited H$\alpha$, H$\beta$, and H$\gamma$ lines in the supernova remnant Tycho, which are substantially broadened by the motion of hot protons. In the UV band, \citet{kras2001} detected prominent \ion{He}{II}, \ion{C}{V}, and \ion{O}{V} CX lines from the comet Hyakutake, and proved that CX is the dominant process for the cometary EUV emission. All these CX emissions are found to be associated with outflows/winds/shocks of ionized matter, which interact with the surrounding cold neutral matter at velocities of a few hundred to a few thousand kilometers per second. As reviewed by \citet{kras2004} and \citet{dennerl2010}, the UV region provides a powerful but as yet not fully exploited probe of CX. So far most of the CX studies using the UV data are restricted to the nearby comets and planets.   

Previously, one difficulty in CX studies in the UV band is the lack of a plasma code. Recently we have published a new atomic model and emission code for CX \citep{gu2016}, which implements the state-of-art theoretical calculations and a few experimental data. It has been successfully applied to several X-ray observations \citep{gu2015, gu2016b, pinto2016}, as well as to a new laboratory measurement \citep{shah2016}. We expect that it can also stimulate new CX research at other wavelengths, in particular the UV band. It is hence non-trivial to verify the new code and the associated atomic data (e.g., wavelengths) before applying it to the observed UV data. 

\begin{figure}[!htbp]
\centering
\resizebox{0.78\hsize}{!}{\includegraphics[angle=0]{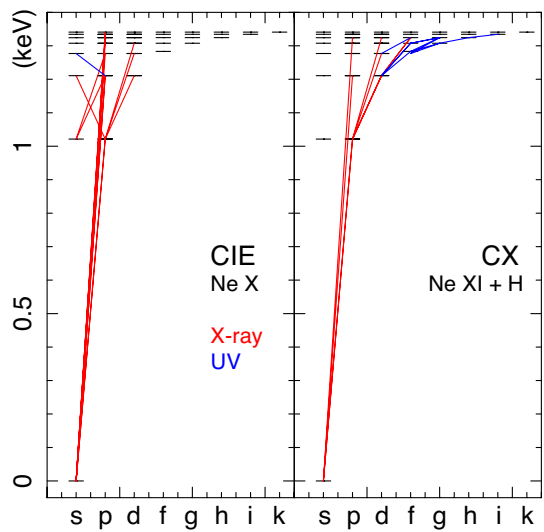}}
\caption{Grotrian diagrams of \ion{Ne}{X} in CIE (left) and CX (right) conditions.
The same balance temperature, kT = 1 keV, is assumed in both cases. The 25 strongest transitions are 
marked. The red and blue lines show
transitions in X-ray and UV bands, respectively. The CX model adopts the atomic-orbital close-coupling calculation in \citet{liu2014}. }
\label{fig:gro}
\end{figure}

\begin{figure}[!htbp]
\centering
\resizebox{0.8\hsize}{!}{\includegraphics[angle=0]{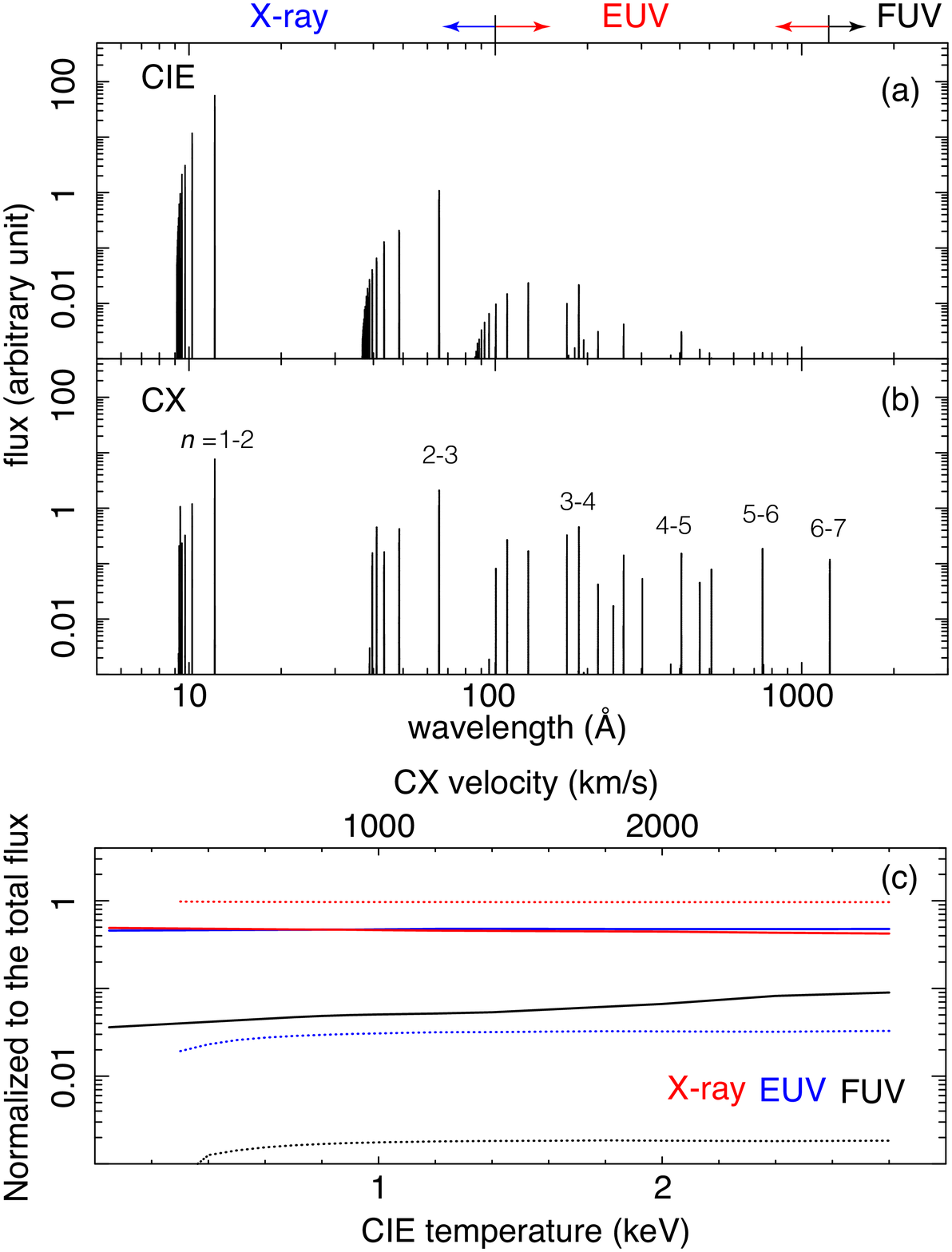}}
\caption{ (a) CIE and (b) CX line spectra of \ion{Ne}{X}. The CIE and CX temperature is set to 1 keV, and the CX velocity is set to 100 km s$^{-1}$. The CX model is based on the AOCC calculation.
(c) FUV (black), EUV (blue) and X-ray (red) parts of the \ion{Ne}{X} emission lines normalized to the total flux, plotted
as a function of temperature and velocity for CIE and CX conditions, respectively. The solid and dotted lines show the CX and the CIE models, respectively.}
\label{fig:uvxray}
\end{figure}

In this paper, we aim to probe CX emission in the UV band from a new potential target, NGC~1275, by making full use of the new CX code. NGC~1275 is the cD galaxy of the Perseus cluster, which is known to host a bright and complex UV source in its core region \citep{baum2005}. The paper is structured as follows: in Sect.~\ref{sect:s2}, we present the calculation of UV lines using the new CX model. Sect.~\ref{sect:s3} describes the analysis and the results from the NGC~1275 UV data. The physical implications of the observed results are presented in Sect.~\ref{sect:s4} and summarized in Sect.~\ref{sect:s5}. Throughout the paper, the errors are given at a 68\% confidence level, and the flux is given in number of photons per second per unit area unless specifically mentioned otherwise.

\section{Charge exchange model \label{sect:s2}}

Charge exchange mostly occurs at the interface of ionized and neutral particles, where ions collide with atoms and capture electrons from the atoms. A unique emission line is produced when the electron, captured into an unstable level with a high principle quantum number $n$, cascades down to the ground state. The CX emission is modeled in a newly developed plasma code as introduced in \citet{gu2015}. Our code calculates the CX line flux by 
\begin{equation}
F = \frac{1}{4 \pi D_{\rm l}^{2}} \int n_{\rm I} n_{\rm N} v \sigma_{\rm I, N}(v,n,l,S) dV,
\end{equation}      
\noindent where $D_{\rm l}$ is the luminosity distance of the object, $n_{\rm I}$ and $n_{\rm N}$
are the densities of ionized and neutral media, respectively, $v$ is the collisional
velocity, $V$ is the interaction volume, and $\sigma_{\rm I, N}$ is the charge exchange cross section, which highly depends on the velocity $v$ and the energy of the capture state characterized by the quantum numbers $n$, $l$, and $S$.

Compared to the line emission excited by free electrons, charge exchange emission shows two main features. First, CX shows characteristic line ratios at X-ray wavelengths. For instance, the transitions from high-$n$ excited levels to the ground are enhanced, and the forbidden-to-resonance line ratios are much larger than under the collisional ionization equilibrium (CIE) condition. This feature has been described in detail in \citet{gu2015,gu2016,gu2016b}. Second, the CX process would produce much more and brighter lines in the UV band than a CIE would, since it introduces many transitions between two highly-excited levels, while a CIE spectrum contains mostly transitions between innershell levels with smaller quantum number $n$ and $l$. As shown in Fig.~\ref{fig:gro}, the SPEX CX model predicts that 15 out of 25 strongest \ion{Ne}{X} CX lines are in the UV band, many of them come
from large $n$ and $l$, e.g., 6h$-$7i, 5g$-$6h, and 4f$-$5g. In contrast, the strong \ion{Ne}{X} CIE lines are mostly (24/25) in X-rays, and the emission is dominated by the Ly$\alpha$ (1s$-$2p) transition. A direct comparison between CIE and CX spectra can be found in Fig.~\ref{fig:uvxray}.

\begin{figure*}[!htbp]
\resizebox{0.95\textwidth}{!}{\includegraphics[angle=0]{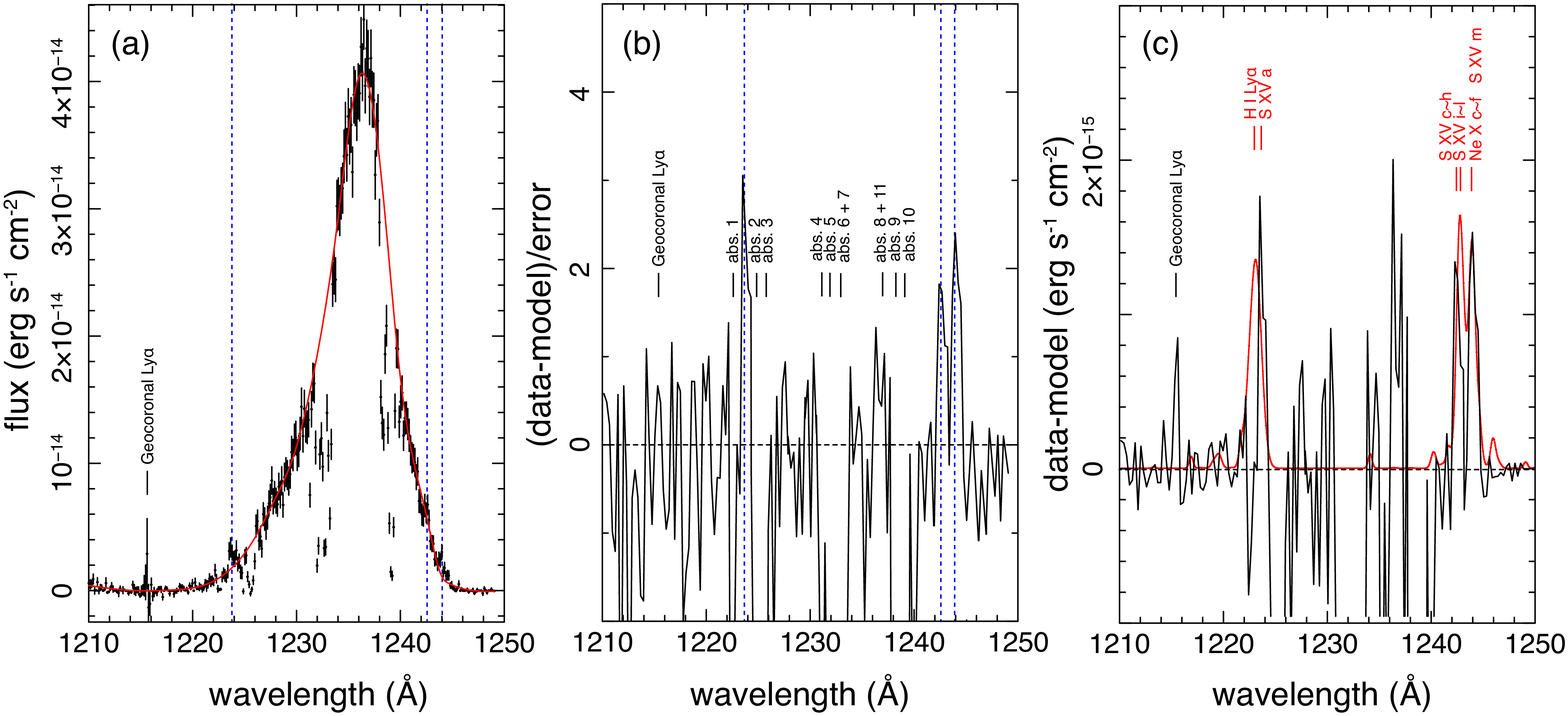}}
\caption{(a) The observed Hubble STIS spectrum of NGC~1275 uncorrected for redenning. Red curve shows the spline fitting to the broadened Ly$\alpha$ line profile. The known features, including the geocoronal line and absorption systems reported in \citet{baum2005} are marked in text. The three unidentified line features are also marked with blue vertical lines. (b) Significance of the emission and absorption features with respect to the spline model. (c) The residual spectrum, plotted together with the CX lines in arbitrary unit. The transitions of the CX lines can be found in Table~\ref{tab:line}.}
\label{fig:spec}
\end{figure*}

We further compare the total CX and CIE fluxes of \ion{Ne}{X} in the FUV (1210~{\AA}$-$2000~{\AA}), EUV (100~{\AA}$-$1210~{\AA}), and X-ray bands (1~{\AA}$-$100~{\AA}), as a function of the ion impacting velocity for CX, and electron temperature for CIE. Based on an atomic-orbital close-coupling (AOCC, \citealt{liu2014}) modelling as shown in Fig.~\ref{fig:uvxray}, the total CX fluxes in the EUV and X-ray bands are comparable, and the FUV emission is about 5 to 15 times lower. Another physical calculation, the multi-channel Landau-Zener (MCLZ, \citealt{mullen2016}) method, predicts comparable X-ray and EUV fluxes as the AOCC ones. The FUV emission based on the MCLZ calculation is a factor of two higher for low speed collisions, meanwhile a factor of eight lower at the high velocity end, than the AOCC value. The uncertainty by the CX physical modellings is discussed in details in Sect.~\ref{sect:inten}. Compared to the CX case, the CIE line spectrum is dominated by X-ray emission, while the EUV and FUV contributions are only about $2-3$\% and $0.1-0.2$\% of the total flux, respectively. The UV versus X-ray lines ratios in both the CX and CIE conditions are relatively insensitive to the variations of velocity or temperature. 

The high UV flux of the CX lines makes the UV band a unique and important window to probe CX emission from astrophysical objects. To further explore the potential of the new CX model, we apply it to an archival UV spectrum as follows.

\section{Data reduction and analysis \label{sect:s3}}

The Hubble STIS observed the NGC~1275 core on Dec 5, 2000. The far-UV Multianode Microchannel Array (MAMA) and the G140M grating were used to cover the redshifted Ly$\alpha$ line. The spectral resolution is 0.08~{\AA}  FWHM, or $\sim 20$ km s$^{-1}$ at the cluster redshift. The observation was carried out in TIME-TAG mode to reject high sky background, and the position angle of the slit was fine-tuned using the quick-look far-UV image to maximize the Ly$\alpha$ brightness. The STIS data are reduced with the standard pipeline using the updated reference files. The same spectrum was reported in \citet{baum2005}. 

As shown in Fig.~\ref{fig:spec}, the STIS spectrum shows a bright broadened Ly$\alpha$ line at about 1237~{\AA}, with many absorption features in the wings. The ten strongest absorption features were reported in \citet{baum2005}. These features have \ion{H}{I} column densities of $10^{12} - 10^{14}$ cm$^{-2}$, velocity shifts of $-3543$ km s$^{-1}$ to $533$ km s$^{-1}$ with respect to the nucleus, and line widths of $15-80$ km s$^{-1}$. \citet{baum2005} suggested that the column densities, shifts, and widths are consistent with the intrinsic absorption features produced by clouds in the nucleus outflow. 

To reveal the additional emission features, it is essential to remove the primary Ly$\alpha$ line profile and any other continuum features. Here we fit these features with a phenomenological ten-knot spline function, and subtract it from the spectrum. The spline-fitting process has been applied in many previous AGN studies using the UV spectra \citep{kraemer2001b, kraemer2001a, shull2012, stevans2014}. As shown in these works, the spline function can provide a reasonable fit to the smoothly varying continuum, as well as the strongly broadened line features. In the fitting, the bands containing the ten strong absorption features are discarded. The ten spline knots are spread evenly across the absorption-free bands, and are adjusted manually to optimize the fitting statistics. The best-fit $\chi^{2}$ is 270 for 231 degrees of freedom. As shown in Fig.~\ref{fig:spec}, the observed Ly$\alpha$ line profile is in general well described by the spline function, except for a few excesses at around 1224~{\AA}, 1237~{\AA}, 1242~{\AA}, and 1244~{\AA}. The positive residual at 1237~{\AA} indicates a more cuspy Ly$\alpha$ profile than the best-fit model. We do not investigate this substructure further as it is not the aim of this paper. To exclude all possible line features, we discarded the four bands that contains excesses, and rerun the spline fit. This gives a best-fit $\chi^{2}$ of 222 for 206 degrees of freedom. The best-fit spline is used as a baseline model for the line detection.  

By varying the spline parameters within their error ranges, we find that the model has a $\sim 4$\% uncertainty at 1210~{\AA}, 3\% at 1224~{\AA}, 2\% at 1237~{\AA}, and 3\% at 1244~{\AA}. To further assess the uncertainty due to the model setting, we test to fit the data with a 15-knot spline function. The best-fit statistics, $\chi^{2}$ of 213 for 201 degrees of freedom, is roughly consistent with that of the baseline model. The new spline deviates by 2\%$-$4\% from the baseline. Hence we consider a systematic uncertainty of 4\%, which is combined with the statistical uncertainty to derive the total error of the data.


\begin{figure}[!htbp]
\resizebox{0.9\hsize}{!}{\includegraphics[angle=0]{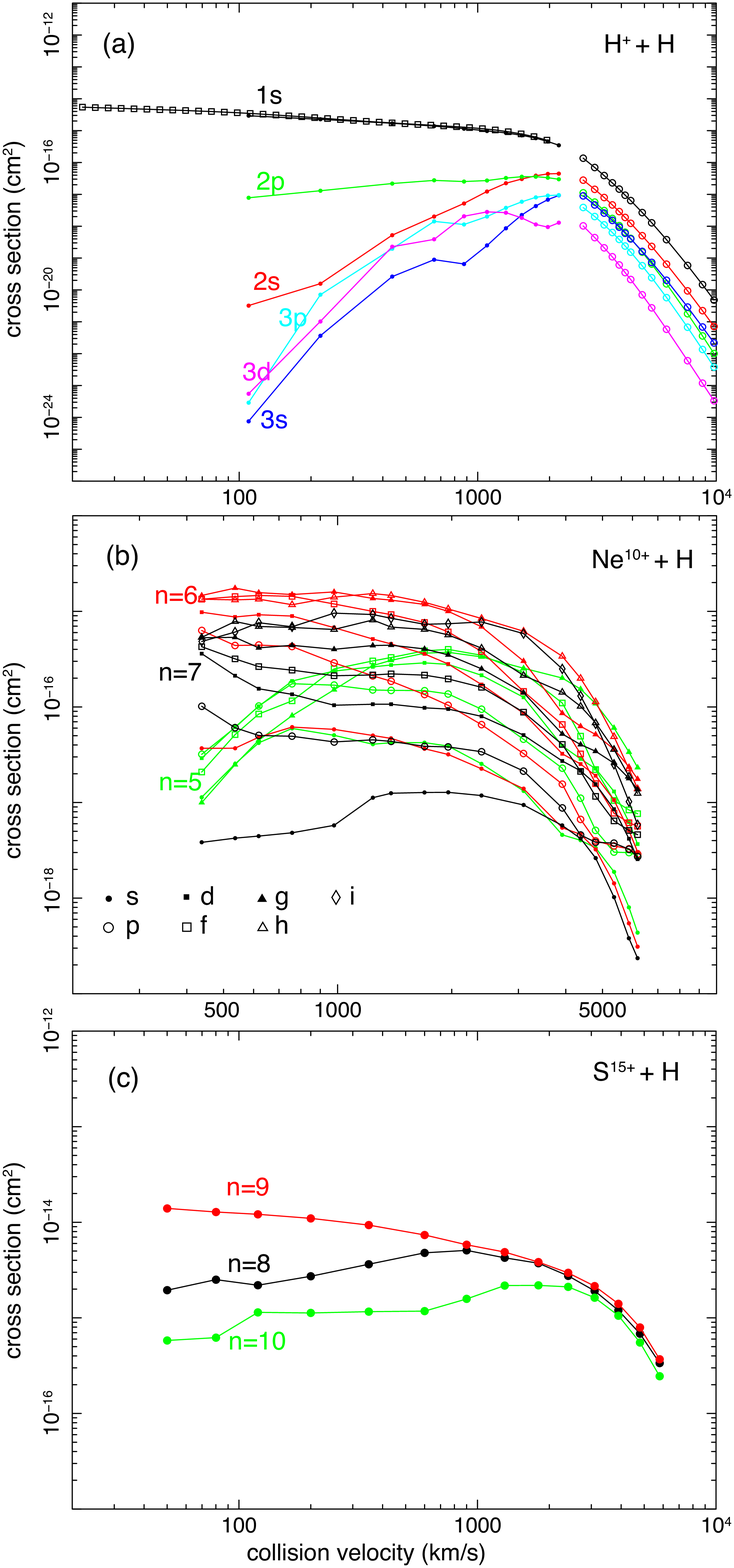}}
\caption{(a) $nl$-resolved CX cross sections of the proton-H atom collision. The data shown by dots, circles and squares are taken from the calculations by \citet{harel1998}, \citet{belkic1992}, and \citet{belkic1991}, respectively. (b \& c) Cross sections of bare Ne 
and H-like S colliding with atomic hydrogen. The Ne and S data are derived from the AOCC method \citep{cumbee2016}, and extrapolation based on data compilation \citep{gu2016}, respectively.}
\label{fig:hcx}
\end{figure}

As plotted in Fig.~\ref{fig:spec}, after subtracting the baseline model, the residual spectrum unveils the line-like features at around 1224~{\AA}, 1242~{\AA}, and 1244~{\AA}, with peak significance of 3.0$\sigma$, 1.8$\sigma$, and 2.3$\sigma$, respectively. The other excesses, including the one at 1237~{\AA}, are all around or below 1$\sigma$. To verify the possible features, we fit the absorption-free band with the baseline spline and three additional Gaussian functions. The new fit gives a $\chi^{2}$ of 240 for degrees of freedom of 222, which is significantly improved from the original fit (270/231). An F-test shows that it is unlikely (null hypothesis probability = 2$\times 10^{-3}$) that the improvement is achieved by chance. The best-fit central energies of the Gaussian components are $1223.6 \pm 0.4$~{\AA}, $1242.4 \pm 0.5$~{\AA}, and $1244.0 \pm 0.2$~{\AA}, which give fluxes of ($1.0 \pm 0.4) \times 10^{-15}$ ergs s$^{-1}$ cm$^{-2}$, ($0.6 \pm 0.3) \times 10^{-15}$ ergs s$^{-1}$ $\rm cm^{-2}$, and ($0.8 \pm 0.3) \times 10^{-15}$ ergs s$^{-1}$ cm$^{-2}$, respectively. The total flux is about 2\% of the primary Ly$\alpha$ feature.

So far the detection is essentially based on a blind line search, in which the probability of a false detection caused by positive random fluctuation must be taken into account (the ``look-elsewhere'' effect; \citealt{gross2010}). Considering that the total bin width of the three weak features are about 1/20 of the absorption-free band, the look-elsewhere effect would boost the null hypothesis probability from 2$\times 10^{-3}$ to 4$\times 10^{-2}$. If these features are indeed identified as atomic lines, we then know the exact line energies, and the look-elsewhere effect can be reduced. Hence we compare these features with the atomic line list as follows.

\subsection{Line identification}

First we search for the target lines in the line list of CIE emission. Since the emission might come from any object, including the cool clouds, the ISM, and the ICM in the central region of NGC~1275, we look into lines in the temperature range of 0.05 keV to 5 keV. Assuming that the clouds might have a line-of-sight velocity of $-3543$ km s$^{-1}$ to $533$ km s$^{-1}$ with respect to the nucleus \citet{baum2005}, we consider a redshift range of 0.00573 to 0.01933, which corresponds to a rest-frame wavelength of 1200~{\AA}$-$1237~{\AA}. The flux threshold is set to be 1\% of the \ion{H}{I} Ly$\alpha$ line at rest-frame 1215.7~{\AA}. In this band, only a few emission lines are present, which are listed in Table~\ref{tab:line} at three representative temperatures. At a low temperature of $0.05$ keV, the \ion{He}{II} and \ion{O}{V} lines appear at 1213~{\AA} to 1218~{\AA}, while at 1 keV, the \ion{Cr}{XX} line can be found at 1205~{\AA}. Most of these lines, except the \ion{He}{II} line at around 1215~{\AA}, disappear at a higher plasma temperature. It is hence unlikely that the observed features at 1223.6~{\AA}, 1242.4~{\AA}, and 1244.0~{\AA} can be explained simultaneously by one thermal emission component, even if the possible line-of-sight velocity is taken into account. Neither can these features be ascribed to CIE emission of local Galactic or geocoronal sources. 

\begin{table*}[!htbp]
\centering
\caption{Related emission lines}
\label{tab:line}
\begin{threeparttable}
\begin{tabular}{lcccccccccccccccc}
\hline
Type & kT(keV)$^{a}$ & name & lower level & upper level & wavelength({\AA}) & fluxes$^{b}$  \\
\hline 
CIE & 0.05 & \ion{He}{II} & 2s $^{2}$S$_{1/2}$  & 4p $^2$P$_{3/2}$    & 1215.095 & 0.013 \\
CIE & 0.05 & \ion{He}{II} & 2p $^{2}$P$_{3/2}$  & 4d $^2$D$_{5/2}$    & 1215.171 & 0.010 \\
CIE & 0.05 & \ion{O}{V}   & 2s$^2$ $^1$S$_0$       & 2s.2p $^3$P$_2$     & 1213.799 & 0.047 \\
CIE & 0.05 & \ion{O}{V}   & 2s$^2$ $^1$S$_0$       & 2s.2p $^3$P$_1$     & 1218.335 & 0.030 \\
\hline
CIE & 1.0  & \ion{He}{II} & 2s $^{2}$S$_{1/2}$  & 4p $^2$P$_{3/2}$    & 1215.095 & 0.010 \\
CIE & 1.0  & \ion{Cr}{XX} & 2s$^2$.2p $^2$P$_{1/2}$& 2s$^2$.2p $^2$P$_{3/2}$& 1205.252 & 0.033 \\
\hline
CIE & 5.0  & \ion{He}{II} & 2s $^{2}$S$_{1/2}$  & 4p $^2$P$_{3/2}$    & 1215.095 & 0.010 \\
\hline
\hline
CX  & 0.05 & \ion{O}{V}   & 2s$^2$ $^1$S$_0$       & 2s.2p $^3$P$_2$     & 1213.799 & 0.329 \\
CX  & 0.05 & \ion{O}{V}   & 2s$^2$ $^1$S$_0$       & 2s.2p $^3$P$_1$     & 1218.335 & 0.207 \\ 
\hline
CX  & 1.0  & \ion{Ne}{X} a& 6f $^2$F$_{5/2}$    & 7g $^2$G$_{7/2}$    & 1236.063 & 0.084 \\
CX  & 1.0  & \ion{Ne}{X} b& 6d $^2$D$_{5/2}$    & 7f $^2$F$_{7/2}$    & 1236.064 & 0.031 \\
CX  & 1.0  & \ion{Ne}{X} c& 6g $^2$G$_{7/2}$    & 7h $^2$H$_{9/2}$    & 1236.278 & 0.109 \\
CX  & 1.0  & \ion{Ne}{X} d& 6f $^2$F$_{7/2}$    & 7g $^2$G$_{9/2}$    & 1236.278 & 0.269 \\
CX  & 1.0  & \ion{Ne}{X} e& 6h $^2$H$_{9/2}$    & 7i $^2$I$_{11/2}$   & 1236.398 & 0.541 \\
CX  & 1.0  & \ion{Ne}{X} f& 6g $^2$G$_{9/2}$    & 7h $^2$H$_{11/2}$   & 1236.398 & 0.330 \\
CX  & 1.0  & \ion{S}{XV} a& 1s.8p $^3$P$_2$     & 1s.9d $^3$D$_3$     & 1216.128 & 0.133 \\
CX  & 1.0  & \ion{S}{XV} b& 1s.8p $^1$P$_1$     & 1s.9d $^1$D$_2$     & 1232.703 & 0.094 \\
CX  & 1.0  & \ion{S}{XV} c& 1s.8g $^3$G$_3$     & 1s.9h $^3$H$_4$     & 1234.939 & 0.124 \\
CX  & 1.0  & \ion{S}{XV} d& 1s.8g $^3$G$_4$     & 1s.9h $^3$H$_5$     & 1234.940 & 0.156 \\
CX  & 1.0  & \ion{S}{XV} e& 1s.8f $^3$F$_2$     & 1s.9g $^3$G$_3$     & 1235.132 & 0.161 \\
CX  & 1.0  & \ion{S}{XV} f& 1s.8f $^3$F$_3$     & 1s.9g $^3$G$_4$     & 1235.152 & 0.217 \\
CX  & 1.0  & \ion{S}{XV} g& 1s.8g $^3$G$_5$     & 1s.9h $^3$H$_6$     & 1235.175 & 0.188 \\
CX  & 1.0  & \ion{S}{XV} h& 1s.8g $^1$G$_4$     & 1s.9h $^1$H$_5$     & 1235.175 & 0.156 \\
CX  & 1.0  & \ion{S}{XV} i& 1s.8h $^3$H$_6$     & 1s.9i $^3$I$_7$     & 1235.341 & 0.081 \\
CX  & 1.0  & \ion{S}{XV} j& 1s.8f $^3$F$_4$     & 1s.9g $^3$G$_5$     & 1235.552 & 0.276 \\
CX  & 1.0  & \ion{S}{XV} k& 1s.8f $^1$F$_3$     & 1s.9g $^1$G$_4$     & 1235.579 & 0.219 \\
CX  & 1.0  & \ion{S}{XV} l& 1s.8d $^3$D$_1$     & 1s.9f $^3$F$_2$     & 1236.102 & 0.119 \\
CX  & 1.0  & \ion{S}{XV} m& 1s.8d $^3$D$_2$     & 1s.9f $^3$F$_3$     & 1236.440 & 0.163 \\
\hline
CX  & 5.0  & \ion{Ne}{X}  & 6f $^2$F$_{7/2}$    & 7g $^2$G$_{9/2}$    & 1236.278 & 0.020 \\
CX  & 5.0  & \ion{Ne}{X}  & 6h $^2$H$_{9/2}$    & 7i $^2$I$_{11/2}$   & 1236.398 & 0.025 \\
CX  & 5.0  & \ion{Ne}{X}  & 6g $^2$G$_{9/2}$    & 7h $^2$H$_{11/2}$   & 1236.398 & 0.040 \\

\hline
\end{tabular}
\begin{tablenotes}
\item[$(a)$] The balance temperature used in the calculation. The CX collision velocity is set to 500 km s$^{-1}$.
\item[$(b)$] Line fluxes normalized to that of the \ion{H}{I} Ly$\alpha$. 
\end{tablenotes}
\end{threeparttable}
\end{table*}

Next we explore the photoionization model. Here we compute a spectrum assuming photoionization equilibrium based on the new pion model in SPEX v3.03. The ionizing radiation from the AGN continuum components is approximated by a power-law model with a photon index of 2, spanning from 0.1 eV to 1 MeV. As shown in \citet{mehdipour2016}, such an ionizing model can roughly reproduce the ionization balance of the photoionized plasma in typical Type 1 AGNs. Then we calculate the spectrum by changing the ionization parameter $\xi$ gradually from 10$^{-2}$ to 10$^{4}$, corresponding to a balance temperature from about 10$^{-3}$ keV to 10 keV. The resulting lines in the band of interest are identical to those of the CIE plasma as listed in Table~\ref{tab:line}. This means that the observed weak features cannot be fully ascribed to a photoionization component.

Finally we investigate the charge exchange model. Similar to the thermal case, we scan the CX line list by changing the ionization state of the projectile ions 
from 0.05 keV to 5 keV. Since the CX emission also varies as a function of collision velocity, we explore a velocity range of 50 km~s$^{-1}$ to 2000 km~s$^{-1}$ to include all the possible transitions. As shown in Table~\ref{tab:line}, the CX model predicts rich emission lines at an ionization temperature of $\sim 1$ keV. The related lines are the cascade emissions of \ion{H}{I}, \ion{Ne}{X}, and \ion{S}{XV} ions. The \ion{H}{I} Ly$\alpha$ transition is created by proton-atom collisions. As seen in Fig.~\ref{fig:hcx}, most of the 
recombined electrons are captured directly into the ground state, as the projectile and target atomic nuclei have essentially the same potential. Then the Ly$\alpha$ line becomes significant only when the collision dynamics is important, i.e., velocity $\geq$ 1000 km s$^{-1}$. The same condition is applied to \ion{He}{II} $n=4$ to $n=2$ transitions at $\sim$1215.1~{\AA} (Table~\ref{tab:line}), which is not yet included in the current CX model in SPEX due to the lack of theoretical and experimental data of the bare He + atomic H collision. It is expected that some dynamical input is required to capture an electron onto $n=4$ of \ion{He}{II}, which has a larger potential than the ground state of a donor hydrogen atom. Hence we treat the \ion{He}{II} 1215.1~{\AA} line as an additional flux of the \ion{H}{I} Ly$\alpha$ line due to their similar properties and proximity in wavelength.

The \ion{Ne}{X} $n=7$ to $n=6$ transitions at $\sim$1236~{\AA} and \ion{S}{XV} $n=9$ to $n=8$ complex at 1216~{\AA}$-$1236~{\AA} are all characteristic CX lines, which are excited by bare Ne and H-like S collisions with neutral atoms, respectively. In Fig.~\ref{fig:hcx} we demonstrate that the upper levels of the related transitions can be populated effectively by the CX. For interactions between bare Ne and atomic hydrogen, most of the captured electrons are expected to fall onto $n=6$ and $n=7$; the two levels take about 70\% and 30\% of the total capture at $v = 100$ km s$^{-1}$, and about 60\% and 30\% at $v=1000$ km s$^{-1}$. 
For H-like sulfur + H charge exchange, the $n=9$ is the dominant level, contributing to about 80\% of the total capture at a few hundred km s$^{-1}$, and at $v=1000$ km s$^{-1}$, the $n=8$ and $n=9$ have a similar share, about 40\% of the total capture (Fig.~\ref{fig:hcx}). The two projectile ions, bare Ne and H-like S require ionization temperatures of $\geq 0.2$ keV and $\geq 0.6$ keV, respectively (see also Fig.~\ref{fig:ion_kt}).

The STIS spectrum does not show the CX features at the systemic velocity of NGC~1275. By shifting the CX lines
within the velocity range of the observed Ly$\alpha$ absorbers, i.e., $-3543$ km s$^{-1}$ to $533$ km s$^{-1}$ with respect to the galaxy core, we find one potential match at a relative velocity of about $-3400$ km s$^{-1}$, as the \ion{H}{I} Ly$\alpha$, \ion{He}{II} n=4 to n=2, and \ion{S}{XV} 1s.8p ($^{3}$P$_{2}$) $-$ 1s.9d ($^{3}$D$_{3}$) transitions would then appear at $1222.7-1223.7$~{\AA}, a complex of \ion{S}{XV} $n=9$ to $n=8$ transitions at $\sim 1242.4$~{\AA} (\ion{S}{XV} c$-$h in Table~\ref{tab:line}), and the \ion{Ne}{X} $n=7$ to $n=6$ lines and one \ion{S}{XV} line at around 1244.0~{\AA} (\ion{Ne}{X} c$-$f and \ion{S}{XV} m in Table~\ref{tab:line}). As shown in Fig.~\ref{fig:spec}, the three putative features can be simultaneously explained by the CX lines emitted at a large line-of-sight velocity relative to the nucleus. 

A remaining issue with this picture is that the peak of the model for the possible feature at 1223.6~{\AA} is slightly offset to shorter 
wavelength with respect to the data (Fig.~\ref{fig:spec}). This might be explained by the fact that the expected relative velocity of the CX emitter is quite close to one of the Ly$\alpha$ absorbers discovered in \citet{baum2005} (No.1 in their table 4), so that the blue wing of the $1222.7-1223.7$~{\AA} CX complex might have been partially absorbed. We will present more insight on the relation between CX emitter and Ly$\alpha$ absorbers in Sect.~\ref{sect:origin}.

Since the CX line energies are not determined by the natural redshift but instead by a scan across the radial velocity range, the look-elsewhere effect cannot be totally eliminated. Taking into account the bin width of the observed weak line features and the assumed velocity range, it would take about seven independent steps to move the CX line template through the absorption-free band of the observed data, therefore the null hypothesis probability obtained in Sect.~\ref{sect:s3} should be increased to 1.4$\times 10^{-2}$ by the look-elsewhere effect.

When shifting the CX model to a line-of-sight velocity of 6595 km s$^{-1}$, i.e., 1334 km s$^{-1}$ higher than that of NGC~1275, we can obtain an alternative potential match of two features. The \ion{H}{I} Ly$\alpha$ and \ion{S}{XV} 1s.8p ($^{3}$P$_{2}$) $-$ 1s.9d ($^{3}$D$_{3}$) lines would be seen at $1242.4-1242.9$~{\AA}, when a \ion{Fe}{XXIV} $n=12$ to $n=11$ CX complex would appear at $\sim 1223.9$~{\AA}. Since the third feature at 1244.0~{\AA} cannot be explained in this setting, it becomes less preferred than the previous one.

The above CX model assumes that each ion experiences one charge exchange along the line-of-sight. In the case of a very dense cloud, an impinging highly ionized particle might undergo multiple CX with the cloud neutrals until it reaches a neutral state. In that case, many transitions of lowly ionized ions would appear in the spectrum. We find that two strong \ion{O}{V} lines, 2s$^2$ ($^{1}$S$_{0}$) $-$ 2s.2p ($^{3}$P$_{2}$) and 2s$^2$ ($^{1}$S$_{0}$) $-$ 2s.2p ($^{3}$P$_{1}$) at 1213.8~{\AA} and 1218.3~{\AA}, would dominate the model spectrum in the UV band. The two lines do not match with any observed features, and hence the present single collision is much more likely than the multiple collision model.

\subsection{Wavelength uncertainties}

\begin{figure}[!htbp]
\resizebox{0.9\hsize}{!}{\includegraphics[angle=0]{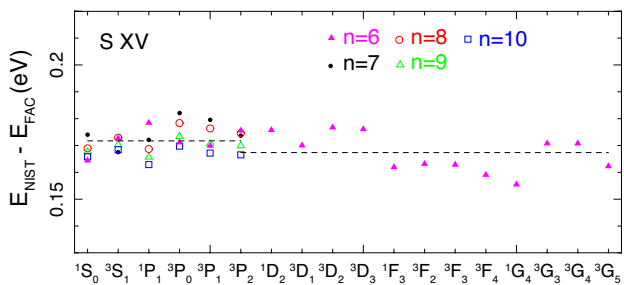}}
\caption{Differences between level energies from NIST database and FAC calculations for $n=6-10$ of \ion{S}{XV}. The two dashed lines show the average offsets for $l=0$ and 1, and for $l=2-4$.}
\label{fig:wav}
\end{figure}

Since the related lines include transitions between two highly-excited levels, it is crucial to examine the uncertainties of the transition energies. The \ion{H}{I} data are taken from the calculated values of \citet{erickson1977}. By comparing with the data compiled in \citet{kramida2010}, we find that the uncertainties are less than 0.00001 eV, and the induced error on the Ly$\alpha$ transition is $\sim 0.001$~{\AA}. The energy level data from \citet{erickson1977} are also used for the \ion{Ne}{X} lines. Here we have crosschecked them with the critically revised data from the latest NIST database \citep{nist5}, and found a $0.0615 \pm 0.0001$ eV energy shift between the two sets. Since this shift is nearly constant for all levels, it will not affect the accuracy of the transitions between two excited levels. Hence the induced wavelength uncertainties are quite small, $\sim 0.01$~{\AA}, for the \ion{Ne}{X} CX lines. 

For \ion{S}{XV}, we use the measured values of \citeauthor{martin1990} (1990; available only for quantum number $l=0$ and 1 in $n \geq 7$) which are present in the latest NIST database,  and derive for the rest levels using the Flexible Atomic Code (FAC; \citealt{gu2008}). The measured values have uncertainties of 0.001 eV. To minimize the deviation between the two data sources, we further calibrate the FAC values to the experimental results. As shown in Fig.~\ref{fig:wav}, the raw FAC calculations for $l=0$ and 1 are systematically smaller than the experimental values by $0.174 \pm 0.001$ eV at $n=8$, and $0.167 \pm 0.001$ eV at $n=9$. For reference, the systematic deviations are $0.172 \pm 0.002$ eV at $n=7$, and $0.167 \pm 0.001$ eV at $n=10$. We also compare the FAC energies of $l=2-4$ of $n=6$ with the measured values, and find a systematic offset of $0.167 \pm 0.003$ eV. Assuming that the FAC values for $l>1$ levels of $n=8$ and 9 share the same average offset as $n=6$, we apply a $nl$-dependent correction to the FAC calculations. The corrected energies agree well with the experimental data within a scatter of about $0.002$ eV and $0.003$ eV for $l\leq1$ and $l>1$, respectively. Hence the final wavelength uncertainties of the \ion{S}{XV} $n=9$ to $n=8$ lines are about 0.2~{\AA} for $l=0$ and 1, and 0.3~{\AA} for levels with larger $l$.

\subsection{Intensity uncertainties \label{sect:inten}}

\begin{figure}[!htbp]
\resizebox{\hsize}{!}{\includegraphics[angle=0]{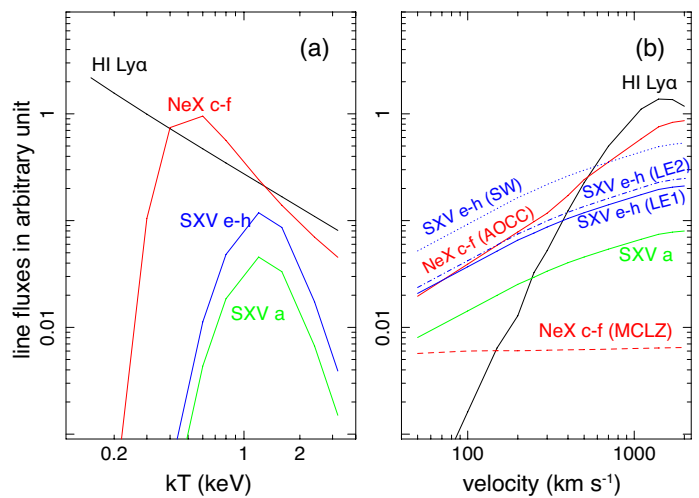}}
\caption{(a) CX line fluxes as a function of ionization temperature of the impinging ions. The detailed transitions can be found in Table~\ref{tab:line}. The collision velocity is set to be 500 km s$^{-1}$. (b) CX line fluxes as a function of collision velocities. The ionization temperature of the projectile ions are set to be 1.2 keV. The solid and dashed red lines show two different calculations of the \ion{Ne}{X} CX. The solid, dotted, and dash-dotted blue lines are the \ion{S}{XV} CX lines assuming $l-$ population functions of low-energy I (LE1), statistical weight (SW), and low-energy II (LE2), respectively (see \citealt{gu2016} for their forms).}
\label{fig:ion_kt}
\end{figure}
In the current model, the relative intensities of the CX lines are rather uncertain, since these lines are highly sensitive to (1) ionization states and abundances of the projectile ions, (2) the collision velocity, (3) the theoretical method for CX calculation and the $l-$ distribution function of the captured electron, and (4) the type of target atoms. None of these properties are known for this object. In Fig.~\ref{fig:ion_kt}, we show the intensities of four representative lines/complexes, i.e., \ion{H}{I} Ly$\alpha$ at 1215.67~{\AA}, the \ion{Ne}{X} $n=7$ to $n=6$ complex at $\sim$1236.3~{\AA}, \ion{S}{XV} 1s.8p ($^{3}$P$_{2}$) $-$ 1s.9d ($^{3}$D$_{3}$) at 1216.13~{\AA}, and the \ion{S}{XV} $n=9$ to $n=8$ complex at $\sim$1235.1~{\AA}, as a function of the ionization temperature. The newly updated ionization balance of \citet{u2017} and the proto-solar abundance of \citet{lodders2009} are utilized in the calculation. For a temperature range from 0.5 keV to 1.0 keV, the intensities of \ion{Ne}{X} and \ion{S}{XV} lines change by a factor of 5 and 250, respectively. As for item (2), we calculate the four sets of lines with the CX velocity changing from 50 km s$^{-1}$ to 2000 km s$^{-1}$. As shown in Fig.~\ref{fig:ion_kt}, the \ion{H}{I}, \ion{Ne}{X}, and two \ion{S}{XV} features have varied dramatically by factors of about 9000, 50, and 10 over the velocity range, respectively. 

The CX model also highly depends on the theoretical method utilized in the rate calculation. As shown in \citet{cumbee2016} and \citet{mullen2016}, the fluxes of high-$n$ transitions differ in general by factors of $2-5$ in different CX calculations. 
For \ion{Ne}{X} $n=7$ to $n=6$ transitions at $\sim$1236.3~{\AA}, the calculations by the AOCC method differs by a factor of 3 to 130 from the values by the MCLZ method (Fig.~\ref{fig:ion_kt}). The large discrepancy is due to the fact that the MCLZ calculation prefers a peak of electron capture at $n=6$, while the AOCC gives a flatter distribution over $n=5$, 6, and 7. On the other hand, the assumed $l-$ distribution of the captured electron will also strongly affect the line fluxes. As shown in Fig.\ref{fig:ion_kt}, the distribution function based on statistical weight gives the largest \ion{S}{XV} line flux, since it can effectively populate the large $l$ shells, which are essentially the upper levels of the 1235.1~{\AA} transitions. The line flux by the more commonly used low energy I weighting function is lower by a factor of $\sim$2.5. Finally, for the item (4), the properties of the target atoms, in particular the ionization potential, might influence the $n$ distribution of the captured electrons (Eq. 3 of \citealt{gu2016}). \citet{mullen2016} showed that the \ion{Fe}{XXVI} CX lines from an atomic He donor can vary by a factor of $\sim 10-100$ relative to a H target.        

As we have shown above, the factors governing the CX line intensities are manifold and highly degenerate. Given the current data, it is not possible to untangle these effects. More observations of CX transitions from the same ions at other wavelengths are apparently needed to resolve the degeneracy.

\section{Discussion \label{sect:s4}}

Based on a new CX model, we prove that the CX process can naturally produce transitions between large $nl$ levels. Many of such transitions fall in the UV range, making it a powerful band to detect and diagnose the astrophysical CX emission.
By re-analyzing the archival Hubble STIS spectrum of the core of NGC~1275 in the Perseus cluster, we identified three weak lines at about 1223.6~{\AA}, 1242.4~{\AA}, and 1244.0~{\AA}, each with a significance of $2-3\sigma$. The putative features are best explained by a CX emission model, with a line-of-sight velocity offset of about $-3400$ km s$^{-1}$ with respect to the NGC~1275 nucleus, which resembles one of the Ly$\alpha$ absorbers reported in \citet{baum2005}. In this model, the lines are emitted when bare hydrogen, neon, and H-like sulfur ions collide with neutral matter. The state-of-art atomic database gives a wavelength uncertainty of $\leq 0.2$~{\AA} for these CX transitions. Due to the model degeneracy, the observed line intensities cannot be used for further interpretation. 

\subsection{Origin of the putative charge exchange component\label{sect:origin}}

First we consider the possibility that the CX emission originates from the hot intracluster medium (ICM) interacting with the H$\alpha$ emission filaments along the line-of-sight of NGC~1275 \citep{conselice2001}.
\citet{walker2015} showed that some of the X-ray emission from the filaments might be indeed CX lines. However,
these filaments have relative velocities of roughly $-300$ to 300 km s$^{-1}$ with respect to the core, 
while our model indicates a relative velocity of $-3400$ km s$^{-1}$. Then the H$\alpha$ filaments cannot be 
the main CX source.

Interactions between the ICM and the cold gas halos of member galaxies might also produce CX emission \citep{gu2013, gu2015}.
Similar interaction is also expected when field galaxies or galaxy groups are merging or infalling into the cluster. 
By searching the SIMBAD database for member galaxies within 5$^{\prime}$ along 
the line-of-sight of NGC~1275, we find that the galaxies cover a relative velocity 
range of $-2400$ km s$^{-1}$ to 3000 km s$^{-1}$, none of them can match with the putative CX component. Thus, the possible CX emission cannot be associated with the member galaxies.

Another possible CX source is the accretion of lowly ionized matter onto the cluster outskirt, which collides
with the hot ions energized by the accretion shocks. However, numerical simulations show that the velocities of the accretion flow are normally $<1000$ km s$^{-1}$ \citep{vazza2011, schaal2016}, which does not match with the observed velocity offset.

One remaining possibility that might apply to the charge exchange component is that it comes from interactions between the AGN outflow gas and the surrounding matter. A similar outflow might be responsible for the absorption features in the wing of the Ly$\alpha$ line reported in \citet{baum2005}. In fact, outflows have been detected extensively in the UV spectra of Seyferts, many of them have similar column densities and velocity offsets as those in NGC~1275 (e.g., \citealt{kriss2000}). \citet{kriss2011} reported that the AGN outflow wind might be in multiple ionization states, which are determined by the balance between time-variable photoionization and recombination. Some clouds might be highly ionized, and they can emit CX emission by interacting with the torus and/or molecular accretion disk, although the observed velocities of the accreting molecular gas in NGC~1275 are $\leq 1000$ km s$^{-1}$ \citep{scharw2013}. In the case of cold outflows, the CX lines would be seen when the wind penetrates into the hot medium surrounding the nucleus. A similar condition has been recently discovered in a Seyfert I AGN NGC~5548 \citep{mao2016}: a CX \ion{N}{VII} X-ray line is detected with the {\it XMM-Newton} RGS, and the CX emitter is likely to be one of the ionized outflowing clouds in that system.

In the outflow scenario, the collision velocity between hot and cold matter should be of the same order as the observed velocity offset ($\sim 3400$ km s$^{-1}$), which is higher than the collision velocity assumed in the current CX model in Table~\ref{tab:line}. At such a large collision velocity, the \ion{H}{I} Ly$\alpha$ and \ion{Ne}{X} lines at 1223.6~{\AA} and 1244.0~{\AA} are expected to become much brighter than the \ion{S}{XV} lines at 1242.4~{\AA} (Fig.~\ref{fig:ion_kt}). However, due to the large uncertainties in both the CX modelling (Sect.~\ref{sect:inten}) and the observed data (Fig.~\ref{fig:spec}), it is practically not possible to derive any useful constraints on the collision velocity for this object.

If the observed emission line features are indeed related to one of the ten outflowing Ly$\alpha$ absorbers reported in \citet{baum2005}, we might expect similar phenomenon from the other absorbers. The No.2 and 3 absorbers in \citet{baum2005} have velocity offsets of about -3000 km s$^{-1}$ and -2800 km s$^{-1}$, respectively. The possible CX lines associated with these two absorbers will be at 1224.6~{\AA} -- 1225.6~{\AA} (\ion{H}{I} and \ion{S}{XV}), 1244.0~{\AA} -- 1245.0~{\AA} (\ion{S}{XV}), and 1245.8~{\AA} -- 1246.6~{\AA} (\ion{Ne}{X} and \ion{S}{XV}). The first two lines might contribute partially to the red wings of the observed features at 1223.6~{\AA} and 1242.4~{\AA}, while the third line is not visible in the current data. It should be noted that the third line might have a larger uncertainty than the first two, since the \ion{Ne}{X} emission is highly model-dependent (Fig.~\ref{fig:ion_kt}). The other absorbers reported in \citet{baum2005} have velocity offsets in the range of $-1407$ km s$^{-1}$ to $533$ km s$^{-1}$, and their possible CX emission lines would either overlap with the absorption dips (\ion{H}{I}), or shift out of the observed energy band (\ion{Ne}{X} and \ion{S}{XV}). Hence it is not surprising that only one or few absorbers can be identified as candidate CX emitters with the current data.

\subsection{CX lines expected at other wavelengths \label{sect:other}}

The current Hubble data can only provide a hint on the possible CX process; further observations covering a broader energy range are required to verify this scenario. Here we discuss the expected CX emission at X-ray and EUV wavelengths based on the CX model presented in Sect.~\ref{sect:s3}. Assuming that the projectile ions have an ionization state of $\sim 1$ keV and a collision velocity of 500 km s$^{-1}$, the brightest CX line of the related ions in the X-ray band would be the \ion{S}{XV} He$\alpha$ forbidden line at a rest-frame energy of 2.43 keV. Further assuming that the \ion{S}{XV} feature at 1242.4~{\AA} has a flux of $0.7 \times 10^{-15}$ erg s$^{-1}$ cm$^{-2}$ (Sect.~\ref{sect:s3}), the flux of the He$\alpha$ CX line is expected to be about $1.1\times10^{-14}$ erg s$^{-1}$ cm$^{-2}$, which is an order of magnitude lower than the thermal emission of the same line based on the CIE modelling of the Perseus ICM reported in \citet{hitomi2016n}. The strongest \ion{Ne}{X} CX line would be the Ly$\alpha$ line at the rest-frame energy of 1.02 keV, which is about 2\% of the CIE line flux. The characteristic \ion{S}{XV} $n=9$ to $n=1$ and \ion{Ne}{X} $n=6$ to $n=1$ transitions would appear at 3.19 keV and 1.32 keV, with fluxes of 1.8$\times 10^{-15}$ erg s$^{-1}$ cm$^{-2}$ and 2.2$\times 10^{-16}$ erg s$^{-1}$ cm$^{-2}$, respectively. 

The brightest CX line in the EUV range is expected to be the \ion{S}{XV} 1s.2s $-$ 1s.2p transition at 673.41~{\AA}, which has a flux of 8.2$\times10^{-15}$ erg s$^{-1}$ cm$^{-2}$. For \ion{Ne}{X}, the 3d $-$ 4f transition at 187.42~{\AA} has a flux of $1.2\times 10^{-15}$ erg s$^{-1}$ cm$^{-2}$.

Assuming solar abundances, we further calculate the CX emission for the other elements. As a notable ion, \ion{O}{VIII} would have a bright CX Ly$\alpha$ of $6.6\times 10^{-14}$ erg s$^{-1}$ cm$^{-2}$, which is more than one order of magnitude lower than the CIE Ly$\alpha$ emission based on the \citet{hitomi2016n} model. The flux of the $n=5$ to $n=1$ transition at 0.84 keV is expected to be $3.6\times 10^{-14}$ erg s$^{-1}$ cm$^{-2}$. Related to the well-known 3.5 keV line issue \citep{gu2015}, this CX component might emit a \ion{S}{XVI} $n=9$ to $n=1$ transition at a rest-frame energy of 3.45 keV, which would have a flux of 5.5$\times 10^{-16}$ erg s$^{-1}$ cm$^{-2}$. This value is substantially lower than the observed \ion{S}{XVI} CX line with the {\it Hitomi} telescope ($\sim 4 \times 10^{-14}$ erg s$^{-1}$ cm$^{-2}$, \citealt{hitomi2016}), suggesting that other CX components might also exist in the Perseus core.

Despite of the low fluxes in X-rays, the expected CX emission might still be detectable, since the CX lines would probably be shifted significantly from the CIE lines of the Perseus cluster. The current {\it Hitomi} data is not suited for this study, due to the lack of collecting area below $\sim 3$ keV \citep{hitomi2016n}. In the future, the high resolution X-ray spectrometers on-board {\it XARM} (X-ray astronomy recovery mission) and {\it Athena} will greatly improve our understanding of the CX phenomenon in this and other extragalactic objects. 

\section{Summary \label{sect:s5}}

Based on a newly developed CX emission model, we demonstrate that the CX process between hot and cold cosmic plasma is naturally more efficient in creating UV emission than the thermal processes. By re-analyzing the Hubble STIS spectrum of the core of NGC~1275, we detect three possible UV line features at 1223.6~{\AA}, 1242.4~{\AA}, and 1244.0~{\AA}, which can be simultaneously explained by CX reactions between highly ionized hydrogen, neon, and sulfur with atomic hydrogen. Our model shows that the CX emitter has a line-of-sight velocity offset of $\sim -3400$ km s$^{-1}$ with respect to NGC~1275, which is roughly consistent with one of the Ly$\alpha$ absorption structures. Hence we speculate that the possible CX emitter and the absorber might share a same origin, the outflowing clouds in the vicinity of the nucleus.

\begin{acknowledgements}

SRON is supported financially by NWO, the Netherlands Organization for
Scientific Research. 

\end{acknowledgements}

\bibliographystyle{aa}
\bibliography{main}

\end{document}